# Lateral quantum confinement regulates charge carrier transfer and biexciton interaction in CdSe/CdSeS core/crown nanoplatelets

Yige Yao,[#,1] Xiaotian Bao,[#,2] Yunke Zhu,[1] Xinyu Sui,[2] An Hu,[1] Peng Bai,[1] Shufeng Wang,[1,3,4,5] Hong Yang,[1,3,4,5] Xinfeng Liu,[*,2,6] and Yunan Gao[*,1,3,4,5]

1. State Key Laboratory for Artificial Microstructure and Mesoscopic Physics, School of Physics, Peking University, Beijing 100871, China

2. CAS Key Laboratory of Standardization and Measurement for Nanotechnology, National Center for Nanoscience and Technology, Beijing 100190, China; University of Chinese Academy of Sciences, Beijing 100049, China

3. Frontiers Science Center for Nano-optoelectronics, Beijing 100871, China

4. Collaborative Innovation Center of Extreme Optics, Shanxi University, Taiyuan 030006, China

5. Peking University Yangtze Delta Institute of Optoelectronics, Nantong 226010, Jiangsu, China

6. Dalian National Laboratory for Clean Energy, Dalian 116023, China

Corresponding authors: gyn@pku.edu.cn, liuxf@nanoctr.cn

ABSTRACT: Charge carrier dynamics essentially determine the performance of various optoelectronic applications of colloidal semiconductor nanocrystals. Among them, two-dimensional nanoplatelets provide new adjustment freedom for their unique core/crown heterostructure. Herein, we demonstrate that by fine-tuning the core size and the lateral quantum confinement, the charge carrier transfer rate from the crown to the core can be varied by one order of magnitude in CdSe/CdSeS core/alloy-crown nanoplatelets. In addition, the transfer can be affected by a carrier blocking mechanism, *i.e.*, the filled carriers hinder further possible transfer. Furthermore, we found that the biexciton interaction is oppositely affected by quantum confinement and electron delocalization, resulting in a non-monotonic variation of the biexciton binding energy with the emission wavelength. This work provides new observations and insights into the charge carrier transfer dynamics and exciton interactions in colloidal nanoplatelets and will promote their further applications in lasing, display, sensing, *etc*.

TOC：

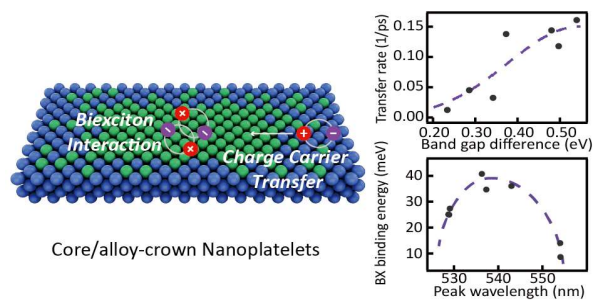

Core/alloy-crown Nanoplatelets

Colloidal two-dimensional nanocrystals have attracted considerable attention over the past decade due to their unique asymmetric geometry, quantum-well energy structure, and many unique optical properties [1-6]. Particularly, cadmium chalcogenide nanoplatelets ($CdX_{(X=S, Se, Te)}$ NPLs) possess favorable optoelectronic properties, such as narrow emission linewidth [1], large exciton binding energy [7, 8], controllable transition dipole moment [9-11], giant oscillator strength [12], and large modal gain coefficients [13], showing great promise in photonic and optoelectronic applications including lasers [14, 15], light-emitting diodes [16], photodetection [17], *etc*. The performance of all these devices is fundamentally determined by various charge carrier dynamics, such as photogeneration, mutual interaction, transfer, and decay, which in turn can be influenced and controlled by the nanocrystal structures.

Heterostructures have been widely used to tune the properties of conventional nanocrystals of quantum dots and rods [18-24]. Fabrication of heterostructures between low-dimensional semiconductors is also a common way to discover novel physical phenomena [25-28]. Two types of heterostructures of NPLs have also been successively developed: core/shell [29] (extension along the surface normal of NPLs) and core/crown [30, 31] (extension along the surface plane). While core/shell heterostructures are very similar in both structure and functionality to their counterparts in quantum dots, core/crown heterostructures are rather unique to NPLs. Unlike the effects of shells that relax strong confinement and lead to large emission shifts, crowns have moderate effects on quantum confinement and emission wavelengths [32]. Therefore, the lateral size and component of the crown can be tuned independently without significantly affecting the core exciton transition. The type-I core/crown heterostructure of CdSe/CdSeS has been exploited in finely tuning the emissions in the blue and green color ranges [33]. Furthermore, the unique core/crown heterostructure enriches the exciton physics, as the core and crown can both absorb and retain charge carriers independently while being adjacently connected [30, 34, 35]. However, the charge carrier and exciton dynamics in core/crown NPLs have not been adequately investigated. The charge carrier transfer efficiency of CdSe/CdS NPLs has been studied [35], but the mechanisms affecting the transfer process are still unclear. Moreover, the studies of the biexciton dynamics were solely carried out in the core-only [36] and core/shell [37, 38] NPLs. One of the reasons is the relatively poor tunability of the ordinary two-step synthesis method of core/crown NPLs [31].

Herein, benefiting from the fine size and band tunability of the newly developed CdSe/CdSeS core/alloy-crown NPLs, we investigate the effect of lateral quantum confinement on the charge carrier transfer dynamics and the biexciton interaction by transient absorption (TA) and time-resolved photoluminescence (TRPL) spectroscopies. The alloyed crown acts as a fine knob to tune the lateral quantum confinement and its consequent influence on the charge carrier and exciton behaviors. We find the transfer rate of the charge carrier from the crown to the core can vary by one order of magnitude for these NPLs with different core sizes and different energy levels. This charge carrier transfer can be also influenced by carrier occupancy under higher photon excitation. For biexcitons, we find their binding energies are affected in opposite ways by quantum confinement and electron delocalization, which has not been observed in NPLs. Overall, this work provides new observations and insights into the charge carrier transfer processes and exciton interactions in core/crown NPLs, which would benefit the development and application of two-dimensional nanocrystals.

RESULTS AND DISCUSSION

**CdSe/CdSeS core/alloy-crown nanoplatelets.** CdSe/CdSeS core/alloy-crown (C/AC) NPLs were synthesized following our previous work [33]. Figure 1(a) shows their geometry and band structure. The CdSe core and CdSeS crown have a gradually alloyed interface, forming the type-I heterostructure with a small conduction band offset and a large valence band offset [31] (also called quasi-type-II heterostructure). Steady-state absorption and PL spectra of a representative 5.5 monolayers (MLs) NPLs sample emitting at 528 nm (labeled as 528-NPLs) are shown in Figure 1(b), where the core gives the absorption and emission peaks at about 528 nm, while the crown leads to a much higher absorbance plateau below 500 nm. The second-order differential of the absorption spectrum (purple line in Figure 1(b)) is used to more precisely determine the transition centers corresponding to its local minimum positions. For this 528-NPLs sample, we can identify two exciton electron-heavy hole (HH) transitions: one at 523 nm for the CdSe core and one at 472 nm for the CdSeS crown. The TA spectrum discussed later confirms this assignment further (Electronic Supplementary Material (ESM) Figure S1(c)).

The alloyed crown not only protects the core from edge defects [39, 40], leading to single-exponential photoluminescence decay (see Figure S1(a)) [41], but also can be used to vary the core size to fine-tune the lateral quantum confinement and the emission wavelengths [42]. The NPLs with smaller core sizes will have stronger lateral quantum confinement and shorter emission wavelengths. Figure 1(c) shows PL spectra of four different 5.5 MLs samples of emission peaks at 554, 545, 536, and 528 nm respectively with decreasing core sizes. We focus on this series of samples in this work. The absorption spectra and transmission electron microscope (TEM) images for different samples are presented in Figure S1(b) and Figure S2.

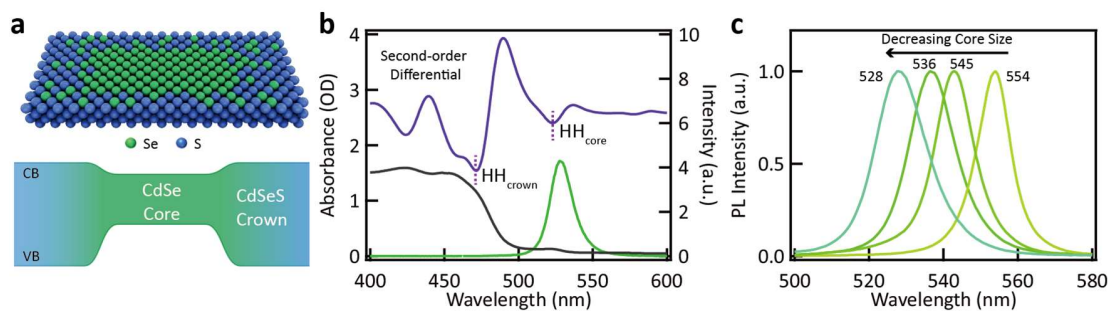

**Figure 1** Basic optical properties of C/AC NPLs. (a) (Upper panel) morphology illustration of the core/alloy-crown NPLs. The green and blue balls represent Se and S atoms respectively, and the Cd atoms are hidden for clarity. (Lower panel) band alignment sketch of C/AC NPLs. (b) PL (green line) and absorption (black line) spectrum of 528-NPLs samples. The second-order differential of the absorption spectrum is also provided (purple line). (c) PL spectrum of different 5.5 MLs C/AC NPLs. The NPLs with shorter emission peaks have smaller core sizes.

**Core and crown exciton emission.** The TRPL spectrum, as shown in Figure 2(a), provides preliminary information on rich charge carrier and exciton dynamics in C/AC NPLs, where three emission features located on different wavelengths can be seen. The prominent long-lived emission can be identified as the core exciton emission directly. The red-side emission feature near the core exciton at this high pump fluence is contributed by the core biexciton emission, which has been reported [14] and is confirmed by the pump fluence-dependent intensity (Figure 2(b)).

The broad blue-side emission feature has not been observed in NPLs, and we attribute it to the crown exciton emission. We note that this emission is absent in core-only CdSe NPLs (Figure S4(a)) and should not be due to emission from higher-order states of the core. Crown emission is consistently observed from low to higher pump fluences (Figure S5), suggesting that these excitons are first photogenerated in the higher energy states of the crown by the 400 nm excitation photons and then directly relax to the band edge of the crown. Therefore, unlike core/shell heterostructure, spatially separated independent excitons are formed in the core and crown respectively. The core exciton emission intensity is saturated at high pump fluence, while the crown exciton emission intensity grows linearly within a larger range (compare green and cyan dots in Figure 2(b)), indicating that the exciton number in the core is limited and the crown can accommodate more excitons. The very short lifetime and weak emission of the crown, in contrast to its higher absorbance, implies rapid charge carrier transfer from the crown to the core.

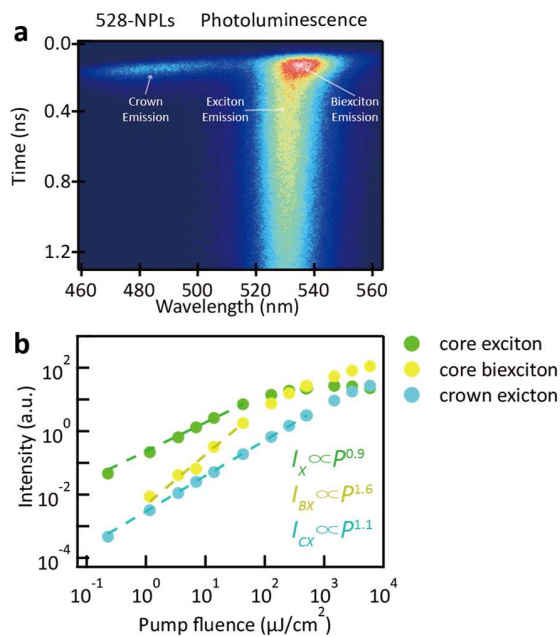

**Figure 2** Core and crown exciton emission in C/AC NPLs. (a) Typical TRPL spectra of 528-NPLs at a high pump fluence. (b) Pump fluence-dependent PL intensity of core exciton (X, green dots), core biexciton (BX, yellow dots), and crown exciton (CX, cyan dots). The PL intensity is obtained

by time-integrating the early 100 ps spectra. The colored dashed lines are the corresponding power function fitting results.

**Charge carrier transfer process at low pump fluence.** To reveal the charge carrier transfer dynamics, we carried out TA spectroscopy measurements with a time resolution of about 120 fs. The TA spectra of 528, 536, and 554-NPLs samples at a low pump fluence (7.1 $\mu$J/cm$^2$) are shown in Figure 3(a-c). The TA spectra clearly show the bleach (negative absorption) of the CdSe core and the CdSeS crown. In Cd-based nanocrystals, it is well established that the TA bleach is mainly caused by the state filling of electrons [36, 43] (above 70% for NPLs [44]), hence the relative absorption change $\Delta A$ is more sensitive to the number of electrons occupied in conduction band. The bleach kinetics of core and crown of 536-NPLs, along the dashed lines in Figure 3(b), are shown in Figure 3(d). The rise of the core bleach coincides well with the decay of the crown, indicating a direct charge carrier transfer from the crown to the core. Similar phenomena are also observed in other C/AC NPLs samples (see Figure S6). The lack of bleach rise in core-only NPLs further confirms this transfer process (Figure S4(b)). The fitting procedure (detailed in ESM section 12) gives a rise time of 29.7 ps for the core and a comparable decay time of 48.9 ps for the crown. We choose the rise time of the core bleach to determine the transfer time because this tens of picoseconds rising comes purely from the charge carrier transfer.

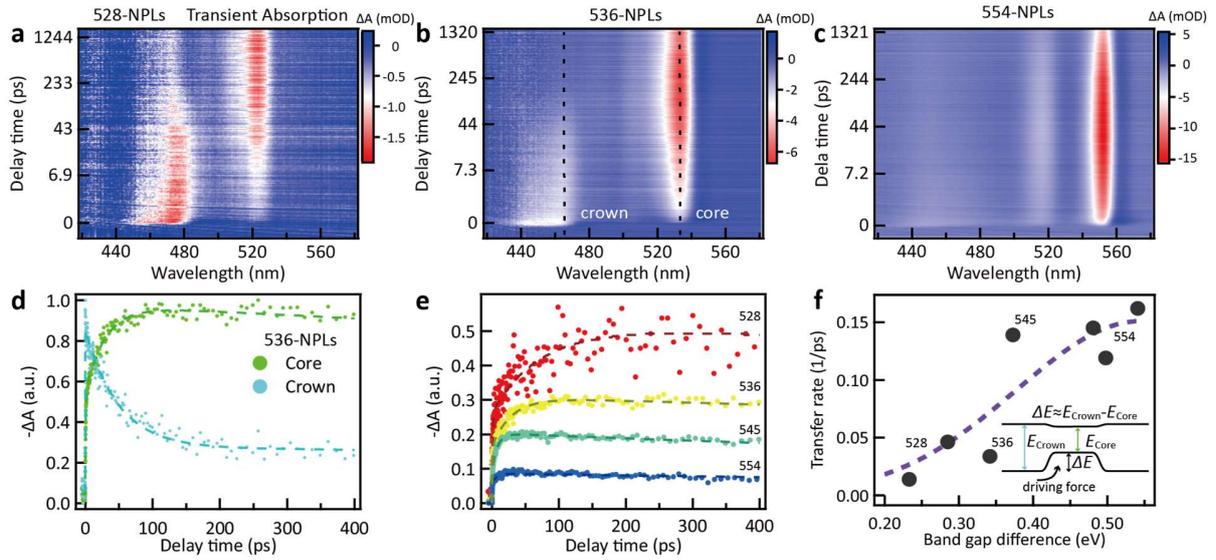

**Figure 3** Charge carrier transfer process of different samples at low pump fluence. (a-c) TA spectra of 528 (a), 536 (b), and 554-NPLs (c) samples at low pump fluence (7.1 μJ/cm²). (d) Core (green dots) and crown (cyan dots) bleach kinetics of 536-NPLs from the dashed line in (b). (e) Core bleach kinetics of 528 (red dots), 536 (yellow dots), 545 (green dots), and 554-NPLs (blue dots). (f) Transfer rate of different samples as a function of band gap difference between the core and crown. The purple dashed line is the fitting result of Marcus theory. The inset is the demonstration of the transfer driving force.

Now we discuss the influence of the core/crown structure on the charge carrier transfer process. The core bleach kinetics of different representative C/AC NPLs samples at a low pump fluence are presented in Figure 3(e). The transfer time shows a large range of change across samples with different emission wavelengths, from nearly a hundred picoseconds for 528-NPLs to several picoseconds for 554-NPLs. Considering the large valence band offset between CdSe and CdSeS, it is widely assumed that the transfer process is dominated by the hole forced by the valence band potential [31, 35]. Therefore, we tend to describe the charge transfer process based on the Marcus theory [45]. The Marcus model has been successfully applied to describe the charge carrier transfer process in heterostructure between quantum dots and other materials [46-49]. In the Marcus model, the charge transfer rate between two materials is stated as:

$$k_{CT} = \frac{2\pi}{\hbar} |H_{a,b}|^2 \frac{1}{\sqrt{4\pi\lambda k_b T}} \exp\left(-\frac{(\lambda - \Delta G)^2}{4\lambda k_b T}\right), \tag{1}$$

where $H_{a,b}$ is the electronic coupling factor between the initial and final states, $\lambda$ is the reorganization energy, $k_b$ is the Boltzmann constant, $\Delta G$ is the driving force (usually the energy difference between the initial and final states). Direct evaluation of the valence band offset is inaccessible, but the band gap difference between the core and crown could be a good approximation because of the very small conduction band offset (inset of Figure 3(f)) [35, 50]. The transfer rate of different samples as a function of the core/crown band gap difference is shown in Figure 3(f). The fitting result of equation (1) is shown as the purple dashed line, and the model describes the data well. Note TA measurement is more sensitive to electrons as mentioned above, what we observed is electron behavior. We infer that charge carrier transfer is driven by the valence band energy difference, while electrons and holes are coupled together through strong Coulomb interactions [51]. Another notable point is that our data don't show the so-called inverted region of the transfer process, *i.e.*, the transfer rate decreases when the driving force exceeds the reorganization energy [47], which is 0.55 eV from our fitting result. This may be due to the limited band gap difference of the samples we can get. There is also the possibility that the Auger-assisted transfer model should be applied [48, 49], but the lack of information about the energy of NPLs' electronic states hinders further confirmation.

It is also possible that the physical dimension rather than the band offset determines the charge carrier transfer process. The transfer behavior of different 528-NPLs samples can resolve this controversy, benefiting from the tunability of the crown size and component. We discuss the comparison of two samples in SI section 7 in detail. Briefly, they both show an emission peak at 528 nm, indicating that they have the same core sizes, while the absorption spectra and TEM images demonstrate that they have different crown band gaps and sizes. If the physical dimension, *i.e.*, mechanisms like charge carrier diffusion or energy transfer determines the transfer process, the NPLs sample with the same core size but the larger crown size would have a smaller transfer rate. However, this is opposite to the actual situation of these two samples (Figure S7), demonstrating that the band gap difference, rather than the physical dimension, is the dominant factor of the transfer process.

Our results demonstrate that the charge carrier transfer process can be regulated by the band gap difference between the core and the crown. As shown in Figure 3, the charge carrier transfer time in the core/crown NPLs can vary by one order of magnitude from several picoseconds to nearly one hundred picoseconds. Compared to the core/shell nanocrystals (at a sub-picosecond level)[52, 53], this transfer (or localization) time is rather longer, partially because of spatially separated excitons in the crown and core. Although the crown is usually believed to act as an exciton funnel [31, 54], this limited charge carrier transfer rate should be paid attention to in the gain application of core/crown NPLs, particularly under femtosecond and picosecond laser-pulse excitations.

**Charge carrier transfer blockade at high pump fluence.** Furthermore, we investigate the bleach kinetics at higher pump fluences, *i.e.*, more charge carriers are generated and mutual interactions may be involved. The TA spectrum of 536-NPLs at a higher pump fluence of 424 $\mu J/cm^2$ is shown in Figure 4(a). Compared to Figure 3(b), the crown bleach has stronger relative intensity and the core bleach reaches its maximum at an earlier delay time. The core bleach kinetics at different pump fluences are shown in Figure 4(b). The rise time of the core bleach decreases rapidly with increasing pump fluence. We attribute this rise-time change to a carrier transfer blockade effect [55], that is, populated charge carriers in the core repel the further transfer from the crown. Figure 4(c) illustrates the carrier transfer without and with blockade under low and high pump fluences. The 400 nm excitation laser will first pump the charge carriers into higher states, then they cool down to the ground state in less than a few picoseconds [36]. At low pump fluence, the core is less likely to be filled because its absorbance is lower than that of the crown, so its bleach rise is mainly contributed by the transfer process of the crown (left part of Figure 4(c)). When the pump fluence is increased, the core will be filled first by the hot carriers due to the fast cooling rate. In this case, further transfer processes are hindered and the rise time reflects more the rapid hot-carrier cooling process (right part of Figure 4(c)).

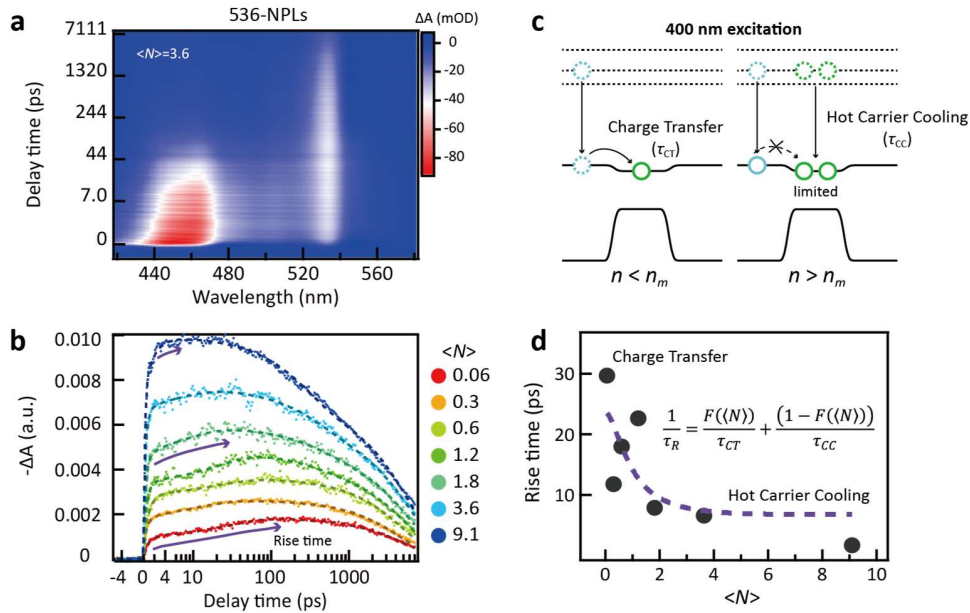

**Figure 4** Charge carrier transfer blockade at higher pump fluences. (a) TA spectrum of 536-NPLs at a high pump fluence (424 μJ/cm$^2$). (b) Core bleach kinetics under different average excited exciton numbers. The purple arrows represent the approximate scale of the bleach rise time. (c)

Illustration of the charge carrier transfer blockade process. (d) The rise time of core bleach as a function of the average excited exciton number. The purple dashed line is a fitting result of the inset equation.

Figure 4(d) shows the rise time as a function of the average excited exciton number $\langle N \rangle$ (calculation of $\langle N \rangle$ is detailed in SI section 8). It shows that with the increasing $\langle N \rangle$ the rise time rapidly decreases to a low level. We quantitatively analyze the change of the rise time based on the Poisson distribution:

$$P_n(\langle N \rangle) = \frac{\langle N \rangle^n e^{-\langle N \rangle}}{n!}. \tag{2}$$

For a given $\langle N \rangle$, the ratio of NPLs having $n$ excitons is $P_n(\langle N \rangle)$. Here we define a saturation exciton number $n_m$. As in Figure 4(c) and the discussion above, for NPLs with $n < n_m$, the rise time is contributed by the transfer process with a lifetime $\tau_{CT}$. While for $n \geq n_m$, the transfer blockade occurs and the rise time is from the hot carrier cooling with a lifetime $\tau_{CC}$. Then the measured rise time for a given $\langle N \rangle$ is determined by the average rate weighted by the Poisson distribution for these two processes:

$$\frac{1}{\tau_R} = \frac{F(\langle N \rangle)}{\tau_{CT}} + \frac{\left(1 - F(\langle N \rangle)\right)}{\tau_{CC}}, \tag{3}$$

$$F(\langle N \rangle) = P_0(\langle N \rangle) + \cdots + P_{n_m-1}(\langle N \rangle). \tag{4}$$

The fitting result of equation (3) for $n_m = 2$ is shown as the purple dashed line in Figure 4(d). Our model describes the trend of the measured rise time change well, supporting this transfer blockade process. Similar procedures are done for other samples, and we find that the best value of $n_m$ is in the range between 2 and 3 for our samples (see Figure S10). Note that because the $\langle N \rangle$ value is determined for the whole NPLs, $n_m$ is also a parameter for the whole. Therefore, the value of $n_m$ is the upper bound of the maximal occupation number in the core. Li *et al.* [56] estimated $n_m = 3.5$ for core-only 4.5 ML CdSe nanoplatelets with lateral area of 167.8 nm$^2$. The relatively smaller values of our results are reasonable as the core sizes should be small to reach the lateral confinement. As a reference, our previous work showed that the core size of C/AC NPLs with emission peak at 550 nm is about 15 nm×8 nm (120 nm$^2$)[33]. Our results show that the core/crown NPLs can regulate the saturation exciton number in the core through lateral confinement, which could be beneficial to the gain application [56].

**Biexciton interaction affected by electron delocalization.** At a high pump fluence, also the biexciton emission of the CdSe core can be recognized. We note that biexciton (BX) interaction is affected by two mechanisms in the core/crown structure. The upper panel of Figure 5(a) shows the TRPL spectrum of the 528-NPLs samples at a high pump fluence of 2990 μJ/cm$^2$. The fast red side emission band relative to the exciton emission band is from the biexciton emission [14] (see the TRPL spectra at different pump fluence in Figure S3). The lower panel of Figure 5(a) gives the spectrum from the upper panel at an early time as indicated by the black dashed line. We use multi-peak fitting (detail in SI section 13) to distinguish the peak information of the exciton and biexciton respectively. For this 528-NPLs sample, the fitting gives exciton and biexciton peaks at 528.9 nm and 534.6 nm respectively. The photon energy difference between them directly yields biexciton binding energy $E_b^{XX} = 25.0$ meV, which is similar to the reported value of NPLs [14].

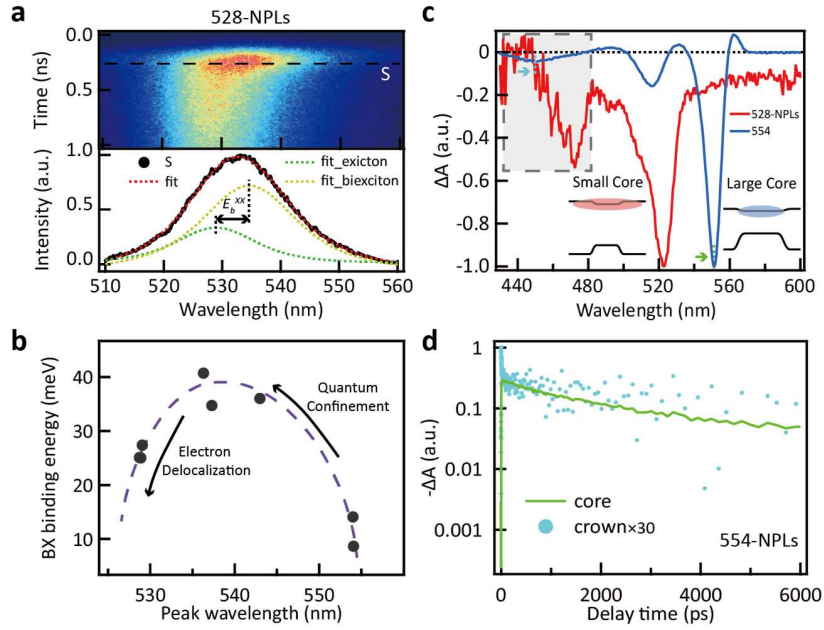

**Figure 5** Effect of core/crown structure on biexciton binding energy. (a) (Upper panel) TRPL spectrum of one of 528-NPLs at high pump fluence (2990 μJ/cm²). (Lower panel) spectrum at an early time and its multi-peak fitting result. (b) Biexciton binding energy as a function of the emission peak wavelength of different samples. The purple dashed line is a guide to the eye. (c) TA spectra of 528 (red line, at 400ps) and 554-NPLs (blue line, at 41ps) after transfer completed. The spectra are normalized at the core bleach peak position. (d) Core (green line) and crown (cyan dots) bleach kinetics of 554-NPLs in a large time range corresponding to the spectral position at the colored dashed line in (c). The crown bleach kinetics is magnified by 30 times.

The biexciton binding energy of different samples is summarized in Figure 5(b) (see other TRPL spectra of different samples in Figure S11), which shows a non-monotonic change as a function of emission peak wavelength. We interpret that this non-monotonic behavior is caused by two mechanisms with opposing effects: quantum confinement and electron delocalization. Previous research had demonstrated that stronger quantum confinement would enhance the biexciton binding energy in semiconductor nanostructures [57-60]. The C/AC NPLs with smaller emission peak wavelengths have smaller core sizes and stronger lateral confinement, which would lead to higher biexciton binding energy. Whereas, the electron delocalization effect, *i.e.*, the electron wave function permeating from the conduction band of the core to the crown due to the small offset, has been deemed to reduce the biexciton binding energy [61-63]. However, whether electrons can be delocalized in CdSe/CdS core/crown NPLs is still debated in the literature [64]. Herein, our TA measurement gives clear evidence that electron does be delocalized in C/AC NPLs. The TA spectra after the transfer completed of 528-NPLs (at 400 ps) and 554-NPLs (at 41ps) are shown in Figure 5(c), which show residual bleach of the crown. This residual crown bleach is strongly sample-dependent as demonstrated by the spectra normalized at the core bleach peak position. One may doubt that the residual bleach originates from the trapped electron/hole in the crown. The time evolution of the crown bleach excludes this speculation. The bleach kinetics of the core and the crown of 554-NPLs in a large time range are shown in Figure 5(d). The residual crown bleach shows the same decay behavior as the core (see other samples in Figure S13), which is not the case with the trapped electron/hole [65]. Therefore, we conclude that this residual crown bleach is contributed by the electron delocalization effect that the core electron wavefunction extends to the crown due to the small conduction band offset and the limited core area [66]. As a result, the NPLs with smaller core sizes and larger crown sizes will have stronger electron delocalization (the inset of Figure 5(c)), causing the decrease of the biexciton binding energy. Our results show that the core/crown heterostructure can affect the core exciton properties through this spatial effect.

CONCLUSION

In summary, we have investigated the influence of the core/crown structure on the charge carrier transfer dynamics and biexciton properties of NPLs by using ultrafast spectroscopies. The charge carrier transfer from the crown to the core can be described by the Marcus charge transfer theory, and the transfer time changes from several picoseconds to nearly one hundred picoseconds for NPLs with different core-crown energy differences. The transfer is additionally affected by carrier blockade due to carrier occupation, which can be quantitatively described by a modal based on the Poisson distribution. We additionally find that the core biexciton binding energy has a non-monotonic change with decreasing emission wavelength due to the combined effect of quantum confinement and electron delocalization. Our findings demonstrate that charge carrier and exciton properties can be largely tuned by nanocrystal heterostructures and provide guidance for further structural design and performance optimization of two-dimensional nanocrystals to achieve the desired applications.

EXPERIMENTAL SECTION

**Synthesis of nanoplatelets**. Synthesis of CdSe/CdSeS core/alloyed-crown nanoplatelets is based on our procedure published earlier [33]. In this work, all the NPLs are 5.5 MLs. Se suspension (Se-Sus) was prepared by dispersing Se powder (79 mg, 1 mmol) in 1-octadecene (ODE) (10 mL) by sonication for 10 min. S suspension (S-Sus) was prepared by dispersing sulfur powder (32 mg, 1 mmol) in ODE (10 mL) by sonication for 30 min. S suspension with oleic acid (OAc) (S-Sus-OAc) was prepared by mixing S-Sus (6 mL) with OAc (200 μL). For core/alloyed-crown nanoplatelets with an emission peak at 550 nm, cadmium myristate (170 mg) in ODE (15 mL) was degassed under vacuum at room temperature for 10 min in a three-necked flask. Then the mixture was heated up to 250°C rapidly under nitrogen. At 250°C, 0.1 m Se (1.5 mL) suspended in ODE was injected. After 60 s, cadmium acetate dihydrate (80 mg) was rapidly added. After 5 min, S-Sus-OAc (1.5 mL) began to inject at 18 mL h$^{-1}$. After the injection was completed, the mixture was cooled down to 110°C by air. OAc (0.3 mL) was injected at 160°C during the cooling process. The whole mixture was degassed for 15 min at 110°C to remove H2S and then cooled to room temperature for purification. By changing the injection time and speed of S (detailed in previous work [33]), one can control the lateral size of the CdSe core and CdSeS crown, resulting in samples with different energy band structures and emission peaks. In general, with a smaller core, the lateral confinement will be stronger and blue-shift the emission peak of the CdSe core. In this way, we can obtain 5.5 ML NPLs with emission peaks continuously ranging between 528 nm and 554 nm. The synthesis of core-only CdSe nanoplatelets with an emission peak at 550 nm was based on the general method [1]. In a three-necked flask, 170 mg of cadmium myristate in 15 mL of ODE was degassed under vacuum at room temperature for 10 minutes. Then the mixture was heated up to 250°C rapidly under nitrogen. At 250 °C 1.5 mL of 0.1 M Se-Sus was injected. After 60 seconds, 120 mg of ground cadmium acetate dihydrate was rapidly added. After 10 minutes, the mixture was cooled down to room temperature by air. 0.15 mL of OAc was injected at 160 °C during the cooling process.

**Time-resolved photoluminescence spectroscopy measurement.** The time-resolved photoluminescence (TRPL) spectroscopy measurements were measured via a streak camera (C10910, HAMAMATSU). Fundamental 800nm pulsed laser (1 kHz, 35 fs) was from a Ti-sapphire amplifier (Legend Elite, Coherent), which was frequency doubled by BBO crystal to obtain 400 nm pump laser. Then the pump laser was focused on the samples through a lens, with a beam area of about 0.01 cm$^2$. The NPLs samples were dispersed in hexane and stored in 10mm quartz cuvettes. During the measurement, the samples were stirred to prevent sample degradation and trion. The emission light is collected from the side and coupled to the streak camera by optical fiber. All the measurements are carried out in room temperature.

**Transient absorption spectroscopy measurement.** The femtosecond transient absorption (TA) spectroscopy measurements were obtained by the HELIOS commercial fs-TA system (Ultrafast Systems). Fundamental 800 nm pulses (1 kHz, 80 fs) from a Coherent Astrella regenerative amplifier were used to pump an optical parametric amplifier (Coherent, OperA Solo) to obtain the frequency-tunable pump beam across the visible region. The pump beam was severed at 500 Hz and focused on the sample with a beam waist diameter of approximately 60 μm. A white light continuum probe beam from 420 to 775 nm was acquired (with a beam waist of 60 μm at the sample) by focusing a small part of the fundamental 800 nm beam on a sapphire window. The magic angle was set for polarization of the pump and probe respectively to avoid the anisotropic effect. All samples in measurements were in hexane solution with 1 mm quartz cuvettes with stirring. Finally, considering the instrument response function of this system, the system had an ultimate temporal resolution of approximately 120 fs. All the measurements are carried out in room temperature.

AUTHOR INFORMATION

**Corresponding Authors**

**Yunan Gao** - *State Key Laboratory for Artificial Microstructure and Mesoscopic Physics, School of Physics, Peking University, Beijing 100871, China; Frontiers Science Center for Nano-optoelectronics, Beijing 100871, China; Collaborative Innovation Center of Extreme Optics, Shanxi University, Taiyuan 030006, Shanxi, China; Peking University Yangtze Delta*

*Institute of Optoelectronics, Nantong 226010, Jiangsu, China;* Email: gyn@pku.edu.cn

**Xinfeng Liu** - *CAS Key Laboratory of Standardization and Measurement for Nanotechnology, National Center for Nanoscience and Technology, Beijing 100190, China; University of Chinese Academy of Sciences, Beijing 100049, China;* Email: liuxf@nanoctr.cn

**Authors**

**Yige Yao** - *State Key Laboratory for Artificial Microstructure and Mesoscopic Physics, School of Physics, Peking University, Beijing 100871, China*

**Xiaotian Bao** - *CAS Key Laboratory of Standardization and Measurement for Nanotechnology, National Center for Nanoscience and Technology, Beijing 100190, China; University of Chinese Academy of Sciences, Beijing 100049, China*

**Yunke Zhu** - *State Key Laboratory for Artificial Microstructure and Mesoscopic Physics, School of Physics, Peking University, Beijing 100871, China*

**Xinyu Sui** - *CAS Key Laboratory of Standardization and Measurement for Nanotechnology, National Center for Nanoscience and Technology, Beijing 100190, China; University of Chinese Academy of Sciences, Beijing 100049, China*

**An Hu** - *State Key Laboratory for Artificial Microstructure and Mesoscopic Physics, School of Physics, Peking University, Beijing 100871, China*

**Peng Bai** - *State Key Laboratory for Artificial Microstructure and Mesoscopic Physics, School of Physics, Peking University, Beijing 100871, China*

**Shufeng Wang -** *State Key Laboratory for Artificial Microstructure and Mesoscopic Physics, School of Physics, Peking University, Beijing 100871, China; Frontiers Science Center for Nano-optoelectronics, Beijing 100871, China; Collaborative Innovation Center of Extreme Optics, Shanxi University, Taiyuan 030006, Shanxi, China; Peking University Yangtze Delta Institute of Optoelectronics, Nantong 226010, Jiangsu, China*

**Hong Yang -** *State Key Laboratory for Artificial Microstructure and Mesoscopic Physics, School of Physics, Peking University, Beijing 100871, China; Frontiers Science Center for Nano-optoelectronics, Beijing 100871, China; Collaborative Innovation Center of Extreme Optics, Shanxi University, Taiyuan 030006, Shanxi, China; Peking University Yangtze Delta Institute of Optoelectronics, Nantong 226010, Jiangsu, China*

**Author Contributions**

[#]Y.Y. and X.B. contributed equally. Y.Y., S.W. and H.Y. performed the TRPL measurement. X.B. and X.S. performed the TA measurement. Y.Z., A.H. and P.B. synthesized NPL samples. Y.G., X.L. supervised the project. Y.Y. and Y.G. analyzed the data and wrote the manuscript. All authors discussed the results and commented on the manuscript.

**Notes**

The authors declare no competing financial interest.

ACKNOWLEDGMENT

This work was supported by the National Natural Science Foundation of China (Grant No. 61875002), the National Key R&D Program of China (Grant No. 2018YFA0306302), and Beijing Natural Science Foundation (Grant No. Z190005). The author acknowledges the support of the Strategic Priority Research Program of Chinese Academy of Sciences (XDB36000000), and the National Natural Science Foundation of China (11874130, 22073022), and the support from the DNL Cooperation Fund, CAS (DNL202016) of Dalian National Laboratory for Clean Energy.

ASSOCIATED CONTENT

**Supporting Information**

The Supporting Information is available free of charge.

Basic optical properties of C/AC NPLs; TEM images of C/AC NPLs; TRPL spectra in core emission region of 528-NPLs at different pump fluence; comparison between core-only and C/AC NPLs; PL spectra in crown emission region of 528-NPLs at different pump fluence; transfer process of different C/AC NPLs samples; comparison of the transfer process of different 528-NPLs samples; estimation of average excited exciton number per NPL; charge carrier transfer blockade modal; TRPL spectra of different C/AC NPLs samples; electron delocalization for different samples; fitting of the TA bleach kinetics; fitting of exciton and biexciton peaks from the TRPL spectrum (PDF)

# Supporting Information

# Lateral quantum confinement regulates charge carrier transfer and biexciton interaction in CdSe/CdSeS core/crown nanoplatelets

Yige Yao,[#,1] Xiaotian Bao,[#,2] Yunke Zhu,[1] Xinyu Sui,[2] An Hu,[1] Peng Bai,[1] Shufeng Wang,[1,3,4,5] Hong Yang,[1,3,4,5] Xinfeng Liu,[*,2,6] and Yunan Gao[*,1,3,4,5]

1. State Key Laboratory for Artificial Microstructure and Mesoscopic Physics, School of Physics, Peking University, Beijing 100871, China

2. CAS Key Laboratory of Standardization and Measurement for Nanotechnology, National Center for Nanoscience and Technology, Beijing 100190, China; University of Chinese Academy of Sciences, Beijing 100049, China

3. Frontiers Science Center for Nano-optoelectronics, Beijing 100871, China

4. Collaborative Innovation Center of Extreme Optics, Shanxi University, Taiyuan 030006, China

5. Peking University Yangtze Delta Institute of Optoelectronics, Nantong 226010, Jiangsu, China

6. Dalian National Laboratory for Clean Energy, Dalian 116023, China

* Corresponding author: gyn@pku.edu.cn, liuxf@nanoctr.cn

# These authors contributed equally to this work.

## 1. Basic optical properties of C/AC NPLs

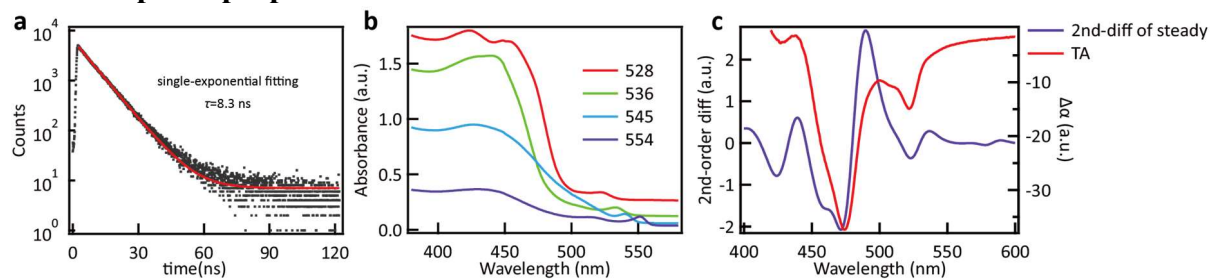

**Figure S1.** (a) TRPL trace of one of the 528-NPLs samples shows single-exponential decay due to the suppression of the traps. (b) Steady-state absorption spectra of different C/AC NPLs samples. (c) Comparison of TA spectrum (red line) and second-order differential of steady-state absorption spectrum (purple line) of 528-NPLs sample. These two spectra show good feature (valley) correspondence.

## 2. TEM images of C/AC NPLs

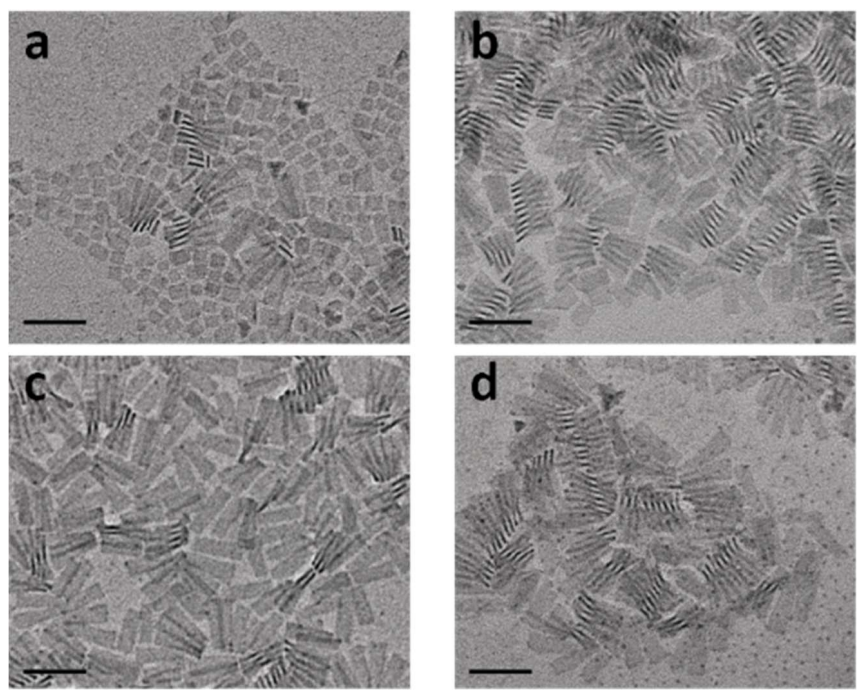

**Figure S2.** The TEM images of 528(a), 536(b), 545(c), and 554-NPLs (d) samples. Scale bars for all panels are 50 nm.

### 3. TRPL spectra in core emission region of 528-NPLs at different pump fluence

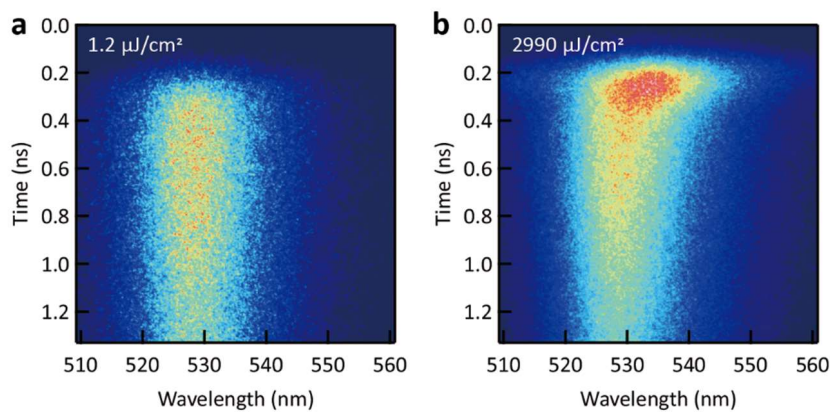

**Figure S3.** TRPL spectra of one of the 528-NPLs samples at low (a) and high (b) pump fluence.

### 4. Comparison between core-only and C/AC NPLs

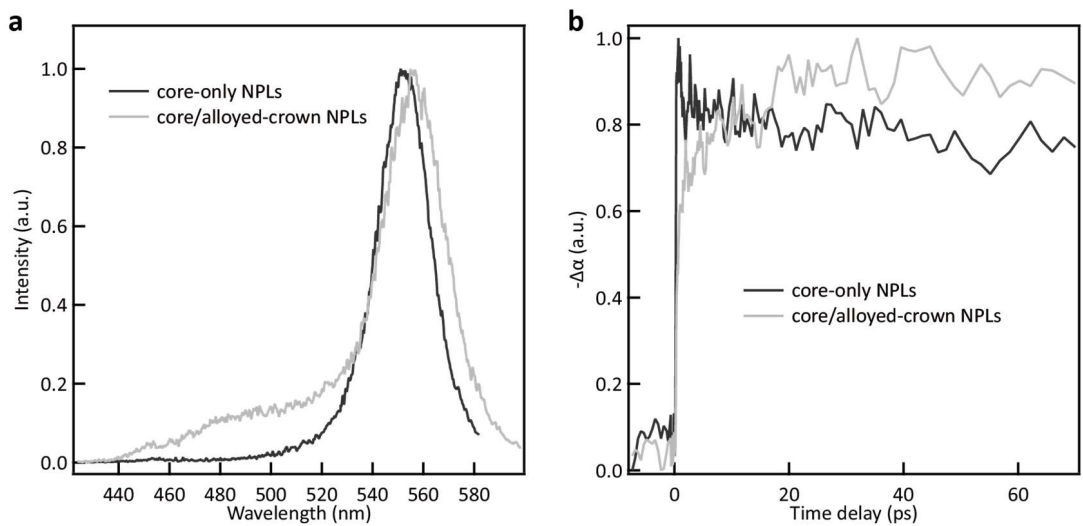

**Figure S4.** (a) Early time-integrated (about 100ps) PL spectra of core-only (black line) and C/AC (gray line) NPLs samples with similar exciton emission peaks (about 550nm) at the same high pump fluence of 3143 uJ/cm$^2$. The spectrum of core-only NPLs clearly shows the lack of the blue-side emission band, confirming that this new band originates from the crown. (b) TA bleach kinetics of core-only (black line) and C/AC (gray line) NPLs samples with similar exciton emission peaks (about 550nm) at the same low pump fluence of 2.08 uJ/cm$^2$. The time evolution of core-only NPLs decays immediately after laser excitation, showing the lack of the transfer process.

## 5. PL spectra in crown emission region of 528-NPLs at different pump fluence

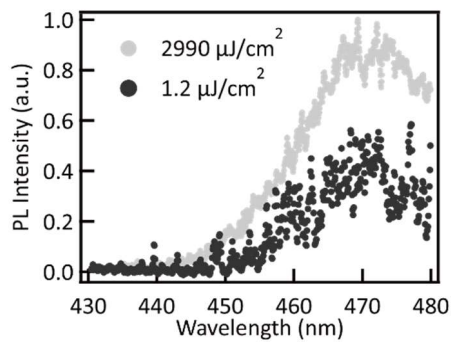

**Figure S5.** PL spectra of one of the 528-NPLs samples in crown emission region at low (black dots) and high (gray dots) pump fluence. Spectra are obtained by time-integrating the first 100ps of TRPL data.

## 6. Transfer process of different C/AC NPLs samples

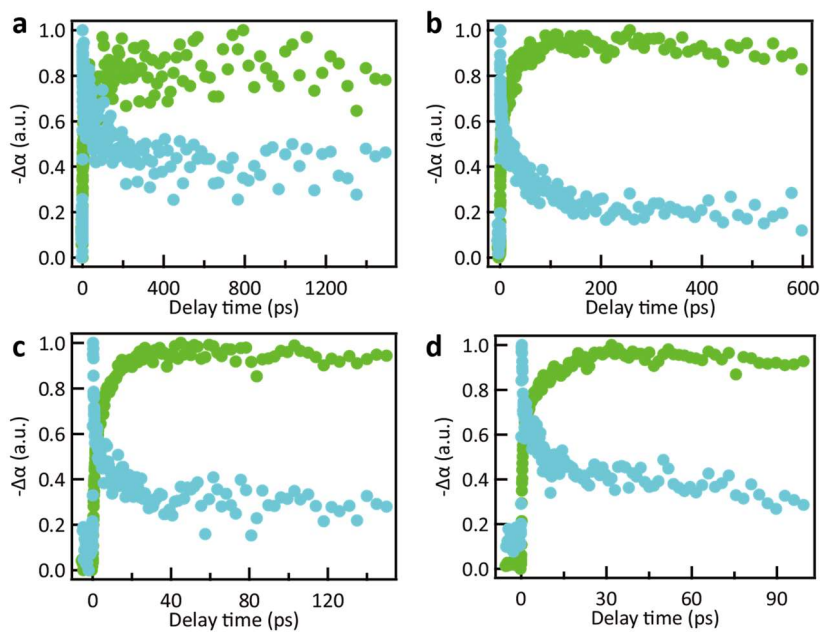

**Figure S6.** TA bleach kinetics of the core (green dots) and crown (blue dots) of 528(a), 536(b), 545(c), and 554-NPLs (d) samples.

## 7. Comparison of the transfer process of different 528-NPLs samples

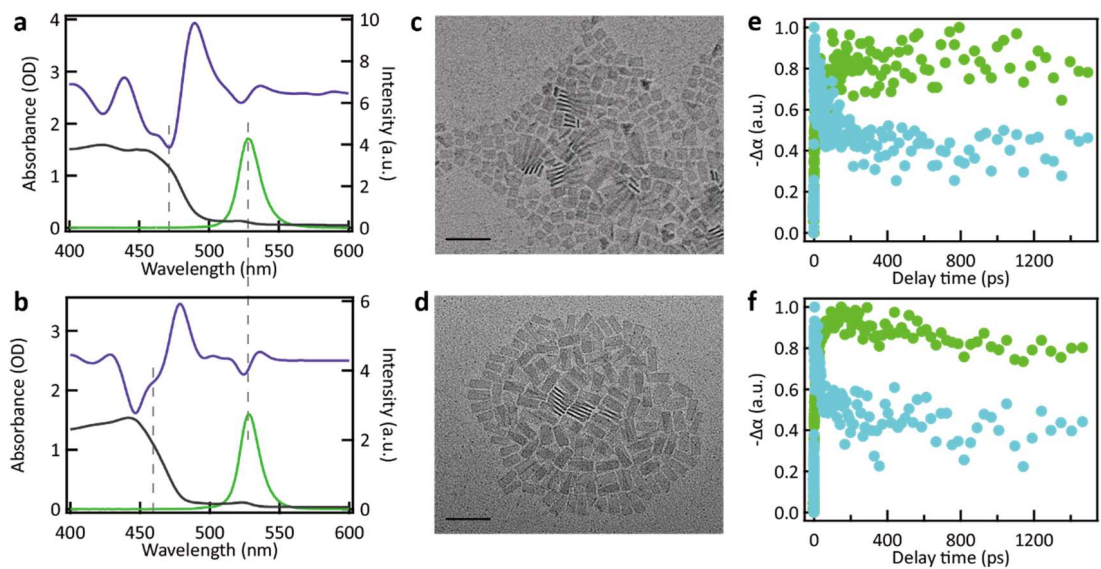

**Figure S7.** Comparison of two 528-NPLs samples. (a-b) Steady-state PL (green lines) and absorption (black lines) spectra of two samples respectively. The purple lines are the second-order differential of the absorption spectra. (c-d) TEM images of these two samples. Scale bars are 50 nm. (e-f) Core (green dots) and crown (cyan dots) bleach kinetics of these two samples.

Figure S7 shows the comparison of two 528-NPLs samples. We label the sample in the first column as 528#1 and that in the second column as 528#2 in the discussion below. The steady-state PL spectra in Figure S7a, b show that they both have an emission peak at 528 nm. Recall that the emission peak is determined by the core size in the core/alloy-crown NPLs, so this indicates that the core sizes of these two samples are the same. On the other hand, the absorption spectra show different features below 500 nm, demonstrating different crown alloy components. The crown transition peak of the 528#1 sample is longer in wavelength, *i.e.*, it has a smaller crown band gap. We derive the band gap difference of the core and the crown as 0.23 eV and 0.29 eV for samples #1 and #2 respectively.

The TEM images in Figure S7c, d directly characterize the sizes of these two samples. We evaluate the average length (width) as 12.5 nm (10.5 nm) and 21.9 nm (9.6 nm) for samples #1 and #2 respectively. Sample #1 is nearly half shorter than sample #2 in total length. Considering that they have the same core sizes, the crown size of sample #2 is therefore at least two times larger than sample#1.

The charge carrier transfer behavior of these two samples is shown in Figure S7e, f. Sample #2 shows a much faster transfer process. The transfer rate is determined as 0.013 ps$^{-1}$ and 0.046 ps$^{-1}$ for samples #1 and #2 respectively. Although sample #1 has much smaller crown sizes, its transfer rate is oppositely smaller than sample #2. The transfer rate change is more reasonable under the Marcus model, where a small change of driving force could result in a large change in transfer rate when far from the reorganization energy. From the discussion above, we exclude the physical dimension as the dominant factor of the transfer process.

## 8. Estimation of average excited exciton number per NPL

We estimate the average excited exciton number per NPL $\langle N \rangle$ using the pump fluence-dependent core exciton kinetics. The estimation method follows Ref. [1]. Briefly, the probability of finding NPLs with n absorbed photons is governed by the Poisson distribution:

$$P_n(\langle N \rangle) = \frac{\langle N \rangle^n e^{-\langle N \rangle}}{n!}. \quad (S1)$$

At a long delay time ($t_l$), all fast processes have completed and only the single exciton remains. Therefore, at a long delay time the TA signal amplitudes are proportional to the number of excited NPLs:

$$-\Delta\alpha(t_l) = \gamma(1 - P_0(\langle N \rangle)) = \gamma(1 - e^{-\langle N \rangle}) = \gamma(1 - e^{-CI}), \quad (S2)$$

where $\gamma$ is a scale factor, I is pump fluence and C is a factor dependent on the absorption cross-section. The fitting result will give the value of C, then the value of $\langle N \rangle$. Figure S8 shows the fitting results for different samples.

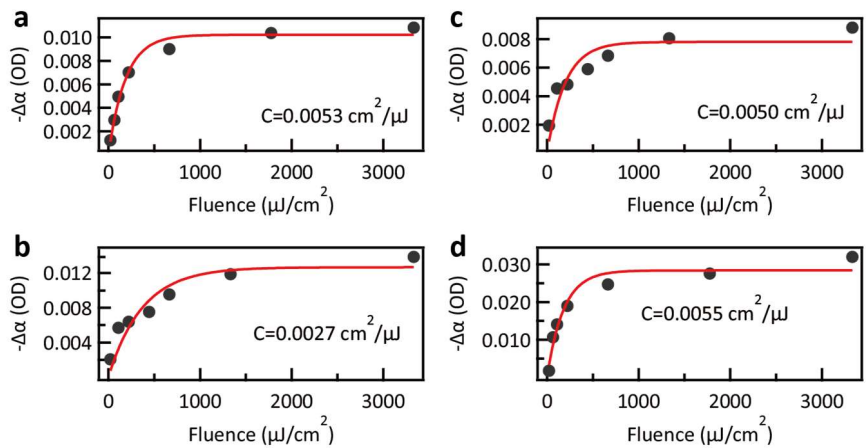

**Figure S8.** Estimation of $\langle N \rangle$ value for 528 (a), 536(b), 545(c), 554-NPLs (d)

## 9. Charge carrier transfer blockade modal

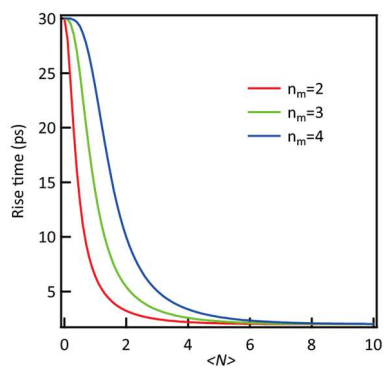

**Figure S9.** Analytical results of eq 3 in the main text

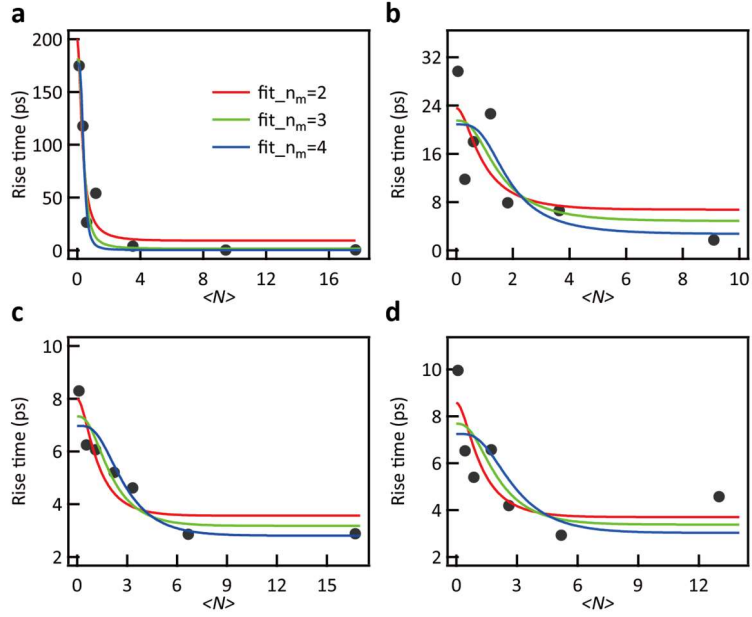

**Figure S10.** Fitting results of eq 3 in the main text of 528 (a), 536 (b), 545 (c), and 554-NPLs (d) samples for $n_m = 2$ (red line), 3 (green line), 4 (blue line).

Figure S9 shows the analytical results of eq 3 in the main text for $n_m = 2,3,4$, where $\tau_{CT} = 30$ ps, $\tau_{CC} = 2$ ps. Figure S10 gives the fitting results of eq 3 in the main text of different C/AC NPLs samples. For larger $n_m$, the fast change of the rise time in small $\langle N \rangle$ value range would be underestimated. On the whole, we conclude that $n_m$ in the range between 2 and 3 are reasonable.

## 10. TRPL spectra of different C/AC NPLs samples

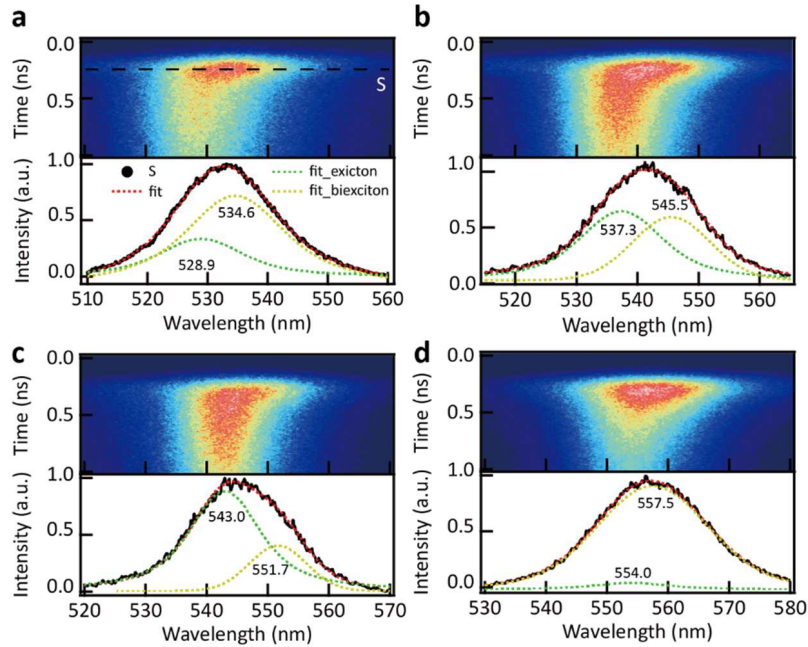

**Figure S11.** TRPL spectra of 528 (a), 536 (b), 545 (c), 554-NPLs (d), and their multi-peak fitting.

## 11. Electron delocalization for different samples

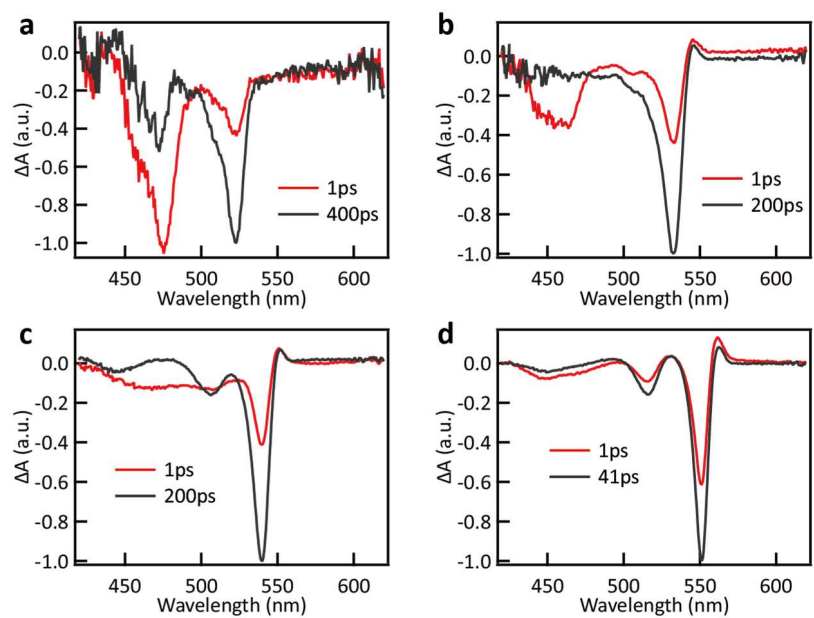

**Figure S12.** The TA spectra of 528 (a), 536 (b), 545 (c), and 554-NPLs (d) samples at short (red line) and long (black line) delay times. The long delay time spectrum of each sample is selected at the time after the transfer process has completed. The spectra are normalized for comparison.

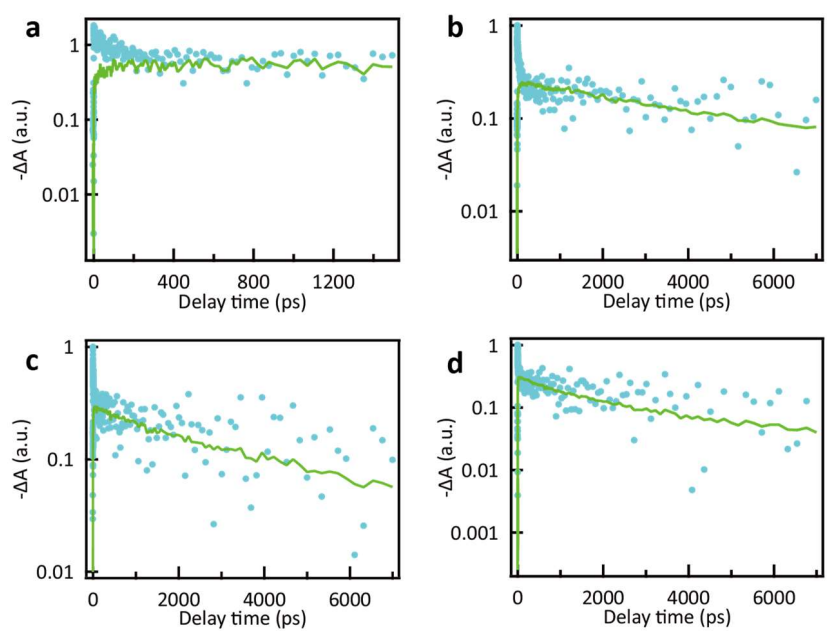

**Figure S13.** The core (green line) and crown (cyan dots) bleach kinetics in the larger delay time range of 528 (a), 536 (b), 545 (c), and 554-NPLs (d) samples. The traces are normalized for clarification. The residual crown bleach all shows the same decay behavior as the core for different samples.

## 12. Fitting of the TA bleach kinetics

We choose the commonly used multi-exponential function convolved with the gaussian function to fit the formation-decay TA bleach kinetics of the core [2, 3]：

$$y(t) = y_0 + \left( \frac{A_1}{\tau_{ff}} \exp\left(-\frac{t}{\tau_{ff}}\right) + \frac{A_2}{\tau_{fs}} \exp\left(-\frac{t}{\tau_{fs}}\right) + \frac{B_1}{\tau_{df}} \exp\left(-\frac{t}{\tau_{df}}\right) + \frac{B_2}{\tau_{ds}} \exp\left(-\frac{t}{\tau_{ds}}\right) \right) * \frac{1}{\sqrt{2\pi}\omega} \exp\left(-\frac{(t-t_c)^2}{2\omega^2}\right), \qquad (S3)$$

where $y_0$ is background constant, $A_i$ and $B_i$ are area, $\tau_i$ is lifetime, $t_c$ and $\omega$ is the center and width of the gaussian function and $*$ is the convolution symbol. The gaussian function acts as the instrument response function (IRF), and $\omega$ is fixed to be 120 fs. $A_i$ are negative and represent the formation process, while $B_i$ are positive and represent the decay process. $\tau_{ff}$ and $\tau_{fs}$ are lifetimes of the fast and slow channels of the formation process respectively. The near or sub-picosecond fast channel relates to the pump excitation and the exciton diffusion in the crown [2, 4]. The over-picosecond level slow channel relates to the charge carrier transfer process and hot carrier relaxation discussed in the main text. $\tau_{df}$ and $\tau_{ds}$ are lifetimes of the fast and slow channel of the decay process respectively. The slow channel is from the intrinsic exciton radiative recombination. The fast channel originates from the trap states at low pump fluence for some samples or the biexciton recombination at high pump fluence.

The formation-decay dynamics represented by the exponential terms in equation S3 are based on the rate equations of the exciton numbers in the core and crown:

$$\frac{dn_{core}}{dt} = -\frac{n_{core}}{\tau_X} + \frac{n_{crown}}{\tau_{CT}}, \qquad (S4)$$

$$\frac{dn_{crown}}{dt} = -\frac{n_{crown}}{\tau_{CX}} - \frac{n_{crown}}{\tau_{CT}}, \qquad (S5)$$

where $n_{core}$ ($n_{crown}$) is the exciton number in the core (crown), $\tau_X$ ($\tau_{CX}$) is the decay lifetime of the core (crown) exciton, and $\tau_{CT}$ is the charge transfer lifetime. Note that we just consider one decay channel here for simplicity and it will not affect the conclusion. The solutions of equation S4, 5 are analytic:

$$n_{crown}(t) = n_{crown}(0) e^{-\left(\frac{1}{\tau_{CX}} + \frac{1}{\tau_{CT}}\right)t}, \qquad (S6)$$

$$n_{core}(t) = n_{core}(0) e^{-\frac{1}{\tau_X}t} + \frac{\frac{\tau_X}{\tau_{CT}} n_{crown}(0)}{\left(\frac{1}{\tau_{CX}} + \frac{1}{\tau_{CT}}\right)\tau_X - 1}\left(e^{-\frac{1}{\tau_X}t} - e^{-\left(\frac{1}{\tau_{CX}} + \frac{1}{\tau_{CT}}\right)t}\right). \qquad (S7)$$

The time evolution of core exciton consists of a formation term $e^{-\left(\frac{1}{\tau_{CX}} + \frac{1}{\tau_{CT}}\right)t}$ with negative amplitude and a decay term $e^{-\frac{1}{\tau_X}t}$ with positive amplitude, which is the origination of $A_i$ and $B_i$ in equation S3.

Table S1 summarizes the fitting results of the charge carrier transfer process of different typical C/AC NPLs. Table S2 lists the pump fluence-dependent fitting results of the typical 536-NPLs sample in the main text Figure 4d.

**Table S1.** Fitting results of charge carrier transfer process of different C/AC NPLs

| Sample / Parameter | 528 | 536 | 545 | 554 |
|---|---|---|---|---|
| $A_1$ (1/ps) | -0.0010±2e-4 | -0.0020±2e-4 | -0.0030±7e-4 | -0.0030±3e-4 |
| $\tau_{ff}$ (ps) | 1.4±0.2 | 0.90±0.11 | 0.64±0.06 | 0.68±0.08 |
| $A_2$ (1/ps) | -0.050±0.009 | -0.090±0.003 | -0.040±0.001 | -0.050±0.002 |
| $\tau_{fs}$ (ps) | 72±11 | 30.0±1.4 | 8.3±0.4 | 10.0±0.5 |
| $B_1$ (1/ps) | NaN | NaN | 3.6±1.9 | 4.8±0.6 |
| $\tau_{df}$ (ps) | NaN | NaN | 1230±280 | 750±50 |
| $B_2$ (1/ps) | 11±3 | 33.0±0.6 | 32±1 | 37.0±0.8 |
| $\tau_{ds}$ (ps) | 8000±2000 | 5340±90 | 5000±700 | 4500±300 |

**Table S2.** Fitting results of the pump fluence-dependent kinetics of typical 536-NPLs sample

| Fluence (μJ/cm²) \ Parameter | 7.1 | 35.4 | 70.7 | 141.5 | 212.2 | 424.4 | 1061 |
|---|---|---|---|---|---|---|---|
| $A_1$ (1/ps) | -0.0020 ±2e-4 | -0.0040 ±3e-4 | -0.0090 ±7e-4 | -0.0060 ±5e-4 | -0.0060 ±1e-4 | -0.0080 ±7e-4 | -0.0001 ±1e-4 |
| $\tau_{ff}$ (ps) | 0.90 ±0.11 | 0.55 ±0.06 | 0.99 ±0.09 | 0.53 ±0.05 | 0.37 ±0.04 | 0.39 ±0.04 | 0.20 ±0.02 |
| $A_2$ (1/ps) | -0.090 ±0.003 | -0.060 ±0.003 | -0.080 ±0.007 | -0.16 ±0.02 | -0.050 ±0.005 | -0.030 ±0.004 | -0.020 ±0.4 |
| $\tau_{fs}$ (ps) | 30.0 ±1.4 | 12.00 ±0.69 | 21.0 ±2.3 | 23.0 ±2.1 | 7.90 ±0.86 | 6.6 ±1.1 | 1.70 ±0.15 |
| $B_1$ (1/ps) | NaN | NaN | NaN | 1.30 ±0.3 | 2.1 ±0.7 | 3.3 ±0.3 | 4.1 ±0.5 |
| $\tau_{df}$ (ps) | NaN | NaN | NaN | 350 ±80 | 390 ±60 | 370 ±30 | 262 ±10 |
| $B_2$ (1/ps) | 330 ±0.6 | 80.0 ±1.1 | 92.0 ±1.6 | 117 ±3 | 146 ±3 | 185 ±4 | 206 ±3 |
| $\tau_{ds}$ (ps) | 5340 ±90 | 4610 ±60 | 5110 ±90 | 5780 ±190 | 6050 ±210 | 6340 ±190 | 6040 ±120 |

### 13. Fitting of exciton and biexciton peaks from the TRPL spectrum

The determination of biexciton binding energy is based on the commonly used multi-peak fitting procedure. In this work, we choose the Voigt line shape, *i.e.*, the convolution of the Gaussian function and Lorentz function, as the peak function:

$$y(x) = y_0 + \left(\frac{2A}{\pi}\frac{w_L}{4(x-x_c)^2+w_L^2}\right) * \left(\sqrt{\frac{4\ln 2}{\pi}}\frac{e^{-\frac{4\ln 2}{w_G^2}x^2}}{w_G}\right) \qquad (S8)$$

where $y_0$ is baseline constant, $A$ is peak area, $x_c$ is peak position, $w_L$ is Lorentz width, $w_G$ is Gaussian width and $*$ is the convolution symbol.

To reduce the ambiguity of the multi-peak fitting, we first fit the exciton emission peak at low pump fluence by eq S8. Then the emission peak at high pump fluence is two-peak fitted, where the peak-shape parameters ($x_c$, $w_L$, $w_G$) of exciton are fixed and only the exciton peak area $A$ is fitted. Table S3 lists the multi-peak fitting results of different C/AC NPLs. Note that the standard deviation of the peak-shape parameters of the exciton is derived from the first fit and distinguished by parentheses.

**Table S3** multi-peak fitting of different typical C/AC NPLs samples

| Sample / Parameter | 528 | 536 | 545 | 554 |
|---|---|---|---|---|
| $A_X$ | 7.8 ±0.7 | 186400 ±1900 | 17.00± 0.08 | 1.2 ±0.2 |
| $x_{cX}$ (nm) | 528.90 (±0.02) | 537.30 (±0.01) | 543.000 (±0.008) | 554.00 (±0.01) |
| $w_{LX}$ (nm) | 11.0 (±0.4) | 10.4 (±0.3) | 10.2 (±0.2) | 10.7 (±0.2) |
| $w_{GX}$ (nm) | 9.5 (±0.3) | 9.6 (±0.3) | 7.1 (±0.2) | 7.5 (±0.2) |
| $A_{XX}$ | 23.9 ±0.8 | 124000 ±3000 | 4.7 ±0.1 | 23.2 ±0.6 |
| $x_{cXX}$ (nm) | 534.6 ±0.2 | 545.50 ±0.06 | 551.70 ±0.04 | 557.50 ±0.03 |
| $w_{LXX}$ (nm) | 16.0 ±0.5 | 0.8 ±0.6 | -0.0 ±0.6 | 4.9 ±0.8 |
| $w_{GXX}$ (nm) | 8.9 ±0.6 | 15.0 ±0.3 | 11.0 ±0.3 | 19.0 ±0.4 |

on charge carrier dynamics and photoluminescence property of CdSe@CdS/ZnS quantum rods. *J. Phys. Chem. C* **2018**, *122*, 6379-6387.